\documentclass[twocolumn,floatfix,superscriptaddress,amsmath,showpacs,showkeys,aps,prb]{revtex4}

\usepackage{t1enc}
\usepackage[final]{graphicx}
\usepackage{graphicx}
\usepackage{epsfig}
\usepackage{bm}

\usepackage{color}
\usepackage[normalem]{ulem}

\usepackage{graphics}

\begin{document}

\title{Finite temperature phase diagram of ultrathin magnetic films without external fields}

\date{\today}

\author{Santiago A. Pighin}
\email{spighin@cab.cnea.gov.ar}
\affiliation{
Centro At\'omico Bariloche, Comisi\'on Nacional de Energ\'{\i}a At\'omica (CNEA, CONICET)\\
Av. E. Bustillo 9500, R8402AGP S. C. de Bariloche, Argentina}
\author{Orlando V. Billoni}
\email{billoni@famaf.unc.edu.ar}
\affiliation{Facultad de
Matem\'atica, Astronom\'{\i}a y F\'{\i}sica, Universidad Nacional
de C\'ordoba, Instituto de F\'{\i}sica Enrique Gaviola (IFEG-CONICET)\\ Ciudad Universitaria, 5000 C\'ordoba, Argentina}						
\author{Sergio A. Cannas}
\email{cannas@famaf.unc.edu.ar}
\affiliation{Facultad de
Matem\'atica, Astronom\'{\i}a y F\'{\i}sica, Universidad Nacional
de C\'ordoba, Instituto de F\'{\i}sica Enrique Gaviola (IFEG-CONICET)\\ Ciudad Universitaria, 5000 C\'ordoba, Argentina}

\begin{abstract}
We analyze the finite temperature phase diagram of ultrathin magnetic films by introducing a mean field theory, valid in the low anisotropy regime, i.e., close to de Spin Reorientation Transition. The theoretical results are compared with Monte Carlo simulations carried out on a microscopic Heisenberg model. Connections between the finite temperature behavior and the ground state properties of the system are established. Several properties of the stripes pattern, such as the presence of canted states, the stripes width variation phenomenon and the associated magnetization profiles are also analyzed.
\end{abstract}

\pacs{75.70.Ak, 05.70.Fh}
\keywords{Ultrathin magnetic films, phase diagram, mean field theory}
%75.70.Ak 	Magnetic properties of monolayers and thin films
%05.70.Fh 	Phase transitions: general studies

\maketitle

\section{Introduction}
\label{Intro}

Despite the increasing growth of knowledge about magnetic  ordering in ultrathin magnetic films during the last decade, both from experimental \cite{PoVaPe2003,WuWoSc2004,WoWuCh2005,PoVaPe2006,Portmann2006,ChWuWoWuScDoOwQi2007,ViSaPoPePo2008,SaLiPo2010,SaRaViPe2010,LiVe2010} and theoretical works\cite{CaStTa2004,MuSt2006,PiCa2007,BaSt2007,ViSaPoPePo2008,PoGoSaBiPeVi2010,PiBiStCa2010,CaCaBiSt2011}, there are still many open questions, specially regarding its finite temperature behavior. One of the main obstacles to advance in these studies is the long range character of the dipolar interactions, which are fundamental to explain pattern formation in those systems. In particular numerical simulations, although have been of great aid\cite{Ja2004,CaMiStTa2006,RaReTa2006,NiSt2007,CaMiStTa2008,CaBiPiCaStTa2008,WhMaDe2008,DiMu2010}, are strongly limited by finite size effects. To avoid them, system
size must be large enough to contain a large number of domains. The main problem relies not on the direct influence of dipolar interactions on the boundary conditions, but on the fact that the basic spatial scale for these systems, namely the typical domain size, scales exponentially with the exchange to dipolar couplings ratio $\delta$ at very low temperatures\cite{Po1998}, and roughly linear with $\delta$ close to the transition to a disordered state\cite{YaGy1988,WuWoSc2004,ViSaPoPePo2008}. Typical values of $\delta$ in ultrathin magnetic films, like Fe based films,  are around\cite{PiBiStCa2010} $\delta \sim 100$, thus implying the necessity of very large system sizes to acommodate a reasonable number of domains.  To perform simulations with  those sizes  represents  up to now, even in the best case (close to the transition), a formidable task. Therefore, knowledge about how the different thermodynamical properties scale with $\delta$ would be very helpful to estimate whether the numerical results for relatively small values of $\delta$ (typically between $3$ and $5$ up to now)  can be extrapolated to more realistic values.

For the analysis of the magnetic properties of ultrathin films, the
out of plane anisotropy to dipolar coupling $\eta$ is also important.
The system behaviour appears to be strongly dependent on experimental
features that modify it, such as the film thickness and the sample
preparation conditions. A strong dependence is also observed in
numerical simulations for\cite{CaBiPiCaStTa2008}  $\delta=3$. For low $\eta$ values, a SRT from a
uniformly magnetized planar phase into a perpendicular striped phase
can happen at finite temperature T (in the absence of an external field), in agreement with a previous
theoretical prediction\cite{PePo1990}. On the other hand, for high values of $\eta$ there is no planar ferromagnetic phase and the system undergoes a direct transition from the striped state into a disordered one. From these numerical results a global $(\eta,T)$ phase diagram was obtained, that is in qualitative agreement with a variety of experimental results\cite{CaBiPiCaStTa2008}. However, for such a small value of $\delta$ certain features can be very different from that expected for large values of $\delta$. For instance, at zero temperature the striped equilibrium state for  $\eta$ above certain critical value $\eta_c$ (where the SRT occurs) is characterized by a stripe width  almost independent of $\eta$ for $\delta< 5$. On the contrary, for values of $\delta\geq 5$, a strong variation of the equilibrium stripe width  with $\eta$ emerges when\cite{PiBiStCa2010} $\eta>\eta_c$. Another feature that depends strongly on the interplay between exchange and anisotropy is the structure of the magnetization pattern close to the SRT. At zero temperature and close to the SRT, the out of plane component of the magnetization presents an almost sinusoidal shape with a large in--plane component, displaying a canted structure\cite{PiBiStCa2010}. For small values of $\delta$ ($\delta<5$) such structure remains for a relatively large interval of values of $\eta$ above the SRT and changes abruptly to a completely perpendicular striped state with sharp domain walls (Ising like state). Consistently, numerical evidences of a canted structure with a sinusoidally shaped  magnetization profile close to the SRT at finite temperature has been recently reported\cite{WhMaDe2008} for $\delta=4.5$. However, as $\delta$ increases the range of anisotropy values at which such canted state is present shrinks at zero temperature\cite{PiBiStCa2010}, becoming almost negligible for realistic values of $\delta$. Hence, it is not clear whether it is expected to be relevant at finite temperature or not.

In this work we analyze the finite temperature phase diagram in the low anisotropy region (close to the SRT) and several related properties using a coarse--grained based mean field  model for ultrathin magnetic films and Monte Carlo simulations on a microscopic model. The main objective of the paper is to discuss which of the observed features of the phase diagram  for low values of $\delta$ are expected to reflect the large $\delta$ behavior. Several properties stripes patterns are also analyzed. The plan of the paper is as follows: in Section \ref{mf} we introduce the coarse grained model and calculate the associated mean field phase diagram.
In Section \ref{mc} we present  Monte Carlo simulations results for a Heisenberg model  and compare them with the previous ones. In Section \ref{conc} we discuss our results.

\section{The Mean Field Model}
\label{mf}

We consider a general phenomenological Landau Ginzburg free energy for  a two dimension ultrathin magnetic film of the form,

\begin{widetext}
\begin{eqnarray}
F[{\bf M}] &=& \frac{1}{2} \int d^2 {\bf x } \left\{ \left( \nabla
{\bf M}({\bf x}) \right)^2 + r_0 {\bf M}^2({\bf x}) + \frac{u}{2}
{\bf M}^4({\bf x}) \right\} + \frac{1}{2\delta} \int d^2 {\bf x} \int
d^2 {\bf x}'\left[ \frac{{\bf M}({\bf x}).{\bf M}({\bf x'})-3 ({\bf n}.{\bf M}({\bf x}))({\bf n}.{\bf M}({\bf x}'))}{\left|{\bf x}-{\bf
x}'\right|^3} \right]\nonumber\\
& & - \frac{\eta}{\delta}  \int d^2 {\bf x } \;M_z^2({\bf x}) \label{Hreal},
\end{eqnarray}
\end{widetext}

\noindent where ${\bf M}= (M_x,M_y,M_z)$ is the coarse grained magnetization, $\delta$ is the exchange to dipolar couplings ratio, $\eta$ is the anisotropy to dipolar coupling ratio and ${\bf n}$ is a unit vector pointing in the ${\bf x}-{\bf
x}'$ direction. A cutoff at some microscopic scale $\Lambda$ is implied in the second integral. We will assume $\Lambda=1$. The temperature dependency comes through $r_0=r_0(T)$.

In order to minimize Eq.(\ref{Hreal}), we propose a variational stripe like solution, i.e., a modulated solution along the $y$ direction, where only Bloch walls between domains are allowed\cite{Po1998}, namely ${\bf M}({\bf x}) ={\bf M}(x)$ and $M_x(x)=0$. We also assume that   modulus of the magnetization is uniform, i.e.

\[
M_y^2(x) + M_z^2(x) = M^2 \;\; \forall\; x.
\]

\noindent This approximation is expected to breakdown for large enough values of $\eta$, where the statistical weight of spin configurations with large in plane components tends to zero, but non uniform out of plane configurations are still expected to minimize the free energy\cite{CaCaBiSt2011}. In fact, in the $\eta\to\infty$ limit the whole effective free energy (\ref{Hreal}) cease to be valid, being replaced by a functional of a scalar order parameter (local out of plane magnetization), without the anisotropy term\cite{CaCaBiSt2011}.

Under the present assumptions, the following form of the dipolar term can be assumed\cite{YaGy1988}

\[
 \frac{L}{\delta} \int d x \int d x' \frac{M_z(x)\, M_z(x')}{\left|{\bf x}-{\bf
x}'\right|^2}
\]

\noindent where we have neglected the self energy term arising from the dipolar energy, since it just implies a constant shift in the anisotropy coefficient $\eta$.

\noindent Then, the variational free energy per unit area reduces to

\begin{widetext}
\begin{equation}
f[{\bf M}] = \frac{1}{2L} \int dx \; \left\{ \left( \frac{\partial M_y}{\partial x}\right)^2 + \left( \frac{\partial M_z}{\partial x}\right)^2 + r_0M^2 + \frac{u}{2}
 M^4 \right\} + \frac{1}{\delta\,L} \int d x \int d x' \frac{M_z(x)\, M_z(x')}{(x-x')^2}
 - \frac{\kappa}{L\delta} \int dx \; \;M_z^2(x) \label{Hreal3}
\end{equation}
\end{widetext}

\noindent where\cite{PiBiStCa2010} $\kappa=\eta-\alpha$ with $\alpha=3.485\ldots$ We can write $M_z(x)=M\, \phi(x)$ where $|\phi(x)| \leq 1$. Then

\begin{equation}\label{fvar}
    f = \frac{1}{2}\,\left(r_0(T) +2\, e/\delta \right)\; M^2 + \frac{u}{4}\; M^4
\end{equation}

\noindent where,

\begin{widetext}
\begin{equation}\label{eevar}
    e[\phi(x)] = \frac{\delta}{2L} \int dx \; \left\{ \left( \frac{\partial \sqrt{1-\phi^2}}{\partial x}\right)^2 + \left( \frac{\partial \phi}{\partial x}\right)^2 \right\}+ \frac{1}{L} \int d x \int d x' \frac{\phi(x)\, \phi(x')}{(x-x')^2}- \frac{\kappa}{L} \int dx \; \;\phi^2(x),
\end{equation}
\end{widetext}

\noindent i.e.  $e[\phi(x)]$ is the microscopic energy per spin of the associated microscopic model (Heisenberg model with out of plane anisotropy, exchange and  dipolar  interactions), for a spin density profile  $(S^x(x),S^y(x),S^z(x))=(0,\sqrt{1-\phi(x)},\phi(x))$. Its minimal energy configuration as a function of the microscopic parameters $(\delta,\eta)$  can be described by variational expressions characterized by different sets of variational parameters $p_1,p_2,\ldots$ that will be described later.
Minimization of the free energy Eq.(\ref{fvar}) leads to:

\begin{eqnarray}
  \frac{\partial f}{\partial M} &=& M \, (r_0 + 2\, e(p_1,p_2,\ldots)/\delta + u\, M^2)=0\label{firstderivM} \\
   \frac{\partial f}{\partial p_i} &=& \frac{M^2}{\delta}\, \frac{\partial e}{\partial p_i}=0 \label{firstderivp}
\end{eqnarray}
We also have;
\begin{equation}\label{2ndderivM}
    \frac{\partial^2 f}{\partial M^2}= r_0 +2\, e/\delta+ 3u\, M^2,
\end{equation}

\begin{equation}\label{2ndderiv1}
     \frac{\partial^2 f}{\partial M \partial p_i} = \frac{2M}{\delta} \, \frac{\partial e}{\partial p_i},
\end{equation}
and
\begin{equation}\label{2ndderiv2}
     \frac{\partial^2 f}{\partial p_i \partial p_j}= \frac{1}{\delta}\, M^2\,\frac{\partial^2 e}{\partial p_i \partial p_j}.
\end{equation}
$M=0$ is always a solution of the extremal equations (\ref{firstderivM})-(\ref{firstderivp}) and the corresponding free energy $f=0$ is independent of the parameter values of $(p_1,p_2,\ldots)$. Hence, all the second derivatives are zero, except

\begin{equation}\label{stabM0}
\left. \frac{\partial^2 f}{\partial M^2}\right|_{M=0} = r_0 + 2\, e/\delta,
\end{equation}

\noindent which controls the stability of the $M=0$ solution.

An ordered solution (local minimum of $f$) with $M \neq 0$ exists whenever  $r_0 +2\, e/\delta<0$. From Eqs.(\ref{firstderivM})-(\ref{firstderivp}) we have that:

\begin{equation}\label{Mmf}
    M^2 = - (r_0 +2\, e/\delta)/u,
\end{equation}
\noindent and
\begin{equation}\label{Mmf2}
     \frac{\partial e}{\partial p_j} = 0.
\end{equation}

\noindent Hence, from Eqs.(\ref{2ndderivM}) and (\ref{Mmf}) it follows that,

\[
\frac{\partial^2 f}{\partial M^2}= 2u\, M^2,
\]

\noindent and from Eqs.(\ref{2ndderiv1}) and (\ref{Mmf2}),
\[
\frac{\partial^2 f}{\partial M \partial p_i} = 0.
\]

\noindent Then, the Hessian matrix of $f$ has a positive eigenvalue $2u\, M^2$ and a diagonal block that equals $ M^2\, H_e/\delta$, where $H_e$ is the Hessian matrix of $e$. Therefore, a local minimum of $f$ has to be a local minimum of $e$.
The free energy of an ordered phase is:

\begin{equation}\label{fordered}
    f = - \frac{1}{4u}\, (r_0(T)+2\, e_{min}/\delta)^2.
\end{equation}

The extremal properties of $e$ are well known\cite{YaGy1988,PiBiStCa2010}.
 For low values of the anisotropy $\eta$ the minimum of $e$ corresponds to a planar ferromagnetic (PF) configuration $\phi(x)=0$. Above certain critical value\cite{YaGy1988,PiBiStCa2010} $\eta_c(\delta)=\alpha+\pi^2/3-\pi^2/2\delta$,  $e$ is minimized by a  striped profile with periodicity $2h$ ($h$ is the stripe width). Close to $\eta_c$ the domain structure corresponds to a canted sinusoidal wall profile (SWP), where $|\phi(x)| = \cos \theta$ is  constant inside the striped domains ($0\leq \theta \leq \pi/2$ is the canting angle) and presents a sinusoidal structured wall of width\cite{YaGy1988} $w$. The energy of the SWP is given by

\begin{widetext}
\begin{equation}\label{eYG}
    e_{SWP}(s,k,\Delta) = \frac{\delta\,k^2}{2\Delta} \;\left(1-\sqrt{1-s^2} \right) + s^2\, \left((\pi^2/3 -\eta) \, (1-\Delta/2) - \frac{\pi\, k}{2}\, G(\Delta) \right),
\end{equation}
\end{widetext}

\noindent where $s = \cos\theta$, $k\equiv \pi/h$, a $\Delta \equiv w/h$, and\cite{YaGy1988}

\begin{equation}\label{GDelta}
    G(\Delta)= \frac{16}{\pi^2}\sum_{m=1,3,\ldots} \frac{1}{m\,(1-m^2 \Delta^2)^2}\, \cos^2\left( \frac{\pi m \Delta}{2}\right).
\end{equation}

\noindent Different approximations of high accuracy for $G(\Delta)$ are available\cite{YaGy1988,WuWoSc2004}, so the values of $(s,k,\Delta)$ that minimize Eq.(\ref{eYG}) can be found numerically for arbitrary values of $(\delta,\eta)$. As the value of $\eta$ is raised above $\eta_c$ the canting angle decreases fast from $\theta=\pi/2$ at $\eta=\eta_c$ to $\theta \approx 0$ and the striped pattern that minimizes $e$ changes to a hyperbolic wall magnetization profile (HPW) whose energy is given by\cite{PiBiStCa2010}

%\begin{widetext}
\begin{equation}\label{eHPW}
    e_{HPW}(k,\Delta)= \gamma\, (1-\Delta/2) + \frac{4\delta}{\pi^2}\, \frac{k^2}{\Delta} -\frac{4k}{\pi}\, \ln \left(\frac{6\pi}{5\Delta} \right),
\end{equation}
%\end{widetext}

\noindent where $\gamma= A-\eta$, with $A=4.5327...$. Eq.(\ref{eHPW}) can be easily minimized\cite{PiBiStCa2010}

If $r_0<0$ and $\eta<\eta_c$ ($e_{min}= 0$),  the global minimum of $f$ corresponds to $M^2 = -r_0/u$ and $s=0$, that is, to a planar ferromagnetic (PF) state with free energy $f= -r_0^2/4u$. When $r_0=0$ and $\eta<\eta_c$ the system undergoes a second order phase transition between the paramagnetic state and the PF one, independently of $\eta$.
We will assume hereafter that $r_0 = a (T-T_F)$, where $T_F$ is the paramagnetic to PF transition temperature.

 \begin{figure}
\begin{center}
\includegraphics[angle=-90,scale=0.36]{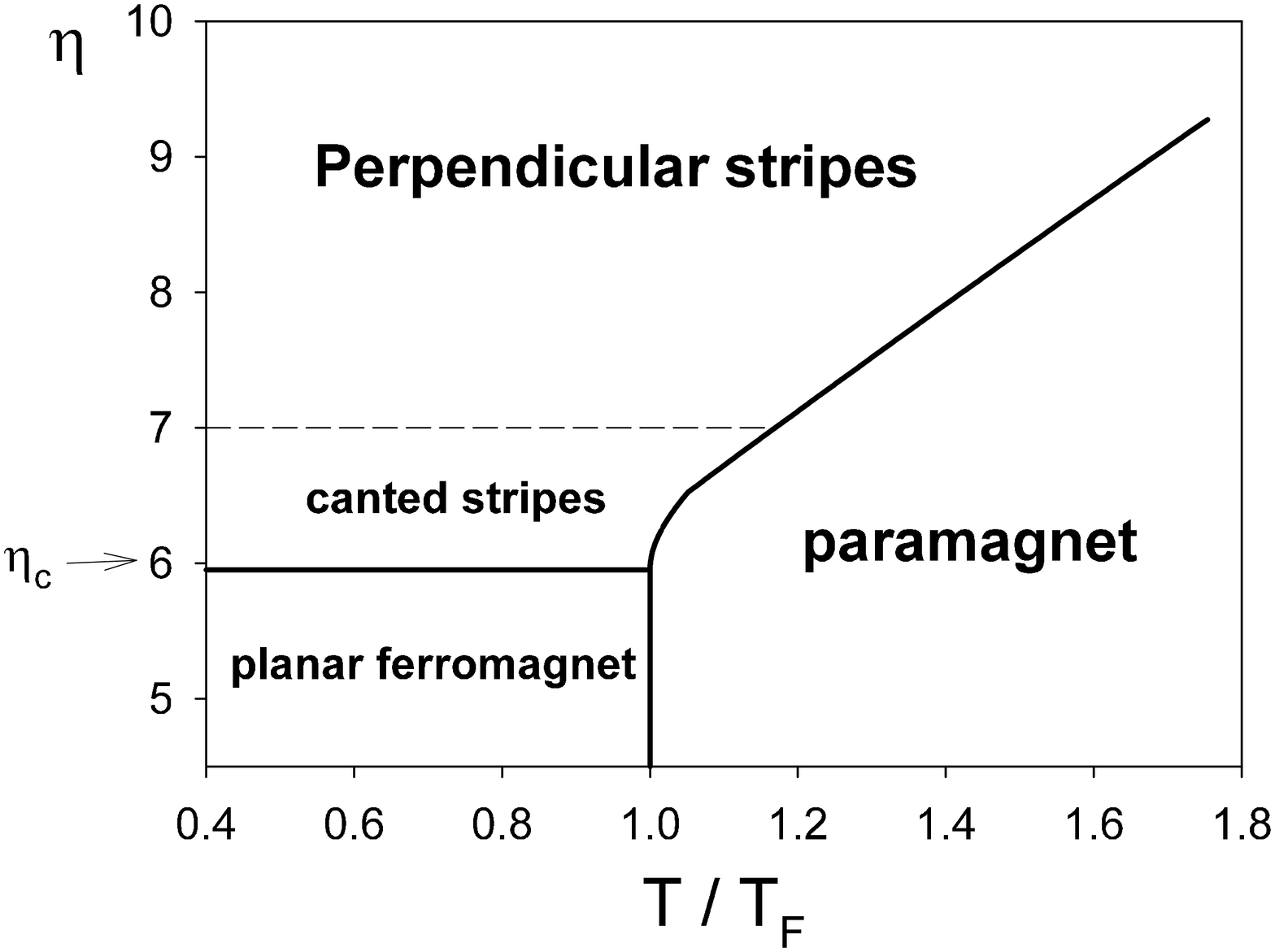}
\caption{\label{pd1} Mean field phase diagram for $\delta=6$ and $a/T_F=1$. Solid lines  correspond to second order phase transitions. The dashed line marks the crossover between canted stripes  and perpendicular stripes configurations; it is estimated arbitrarily as the region above which the maximum in-plane component of the magnetization is less than $5\%$ of the saturation magnetization $M$ ($s < 0.05$)}
\end{center}
\end{figure}

 When $T<T_F$ and $\eta \geq \eta_c$ the  SWP configuration  with free energy given by Eq.(\ref{fordered}) is the stable solution for values of $\eta$ close to $\eta_c$. Since the striped order emerges continuously, the SRT at $\eta=\eta_c$, according to the present approximation, is a second order one. As $\eta$ is further increased there is an energy crossing at certain value of $\eta$ and the stable configuration changes into an HWP. Hence, for $\eta_c\leq \eta\leq \eta^*$ the stable configuration is a canted striped one, while for $\eta > \eta^*$ the stripes are fully saturated in the out of plane direction inside the domains (we will call this state an "Ising striped configuration").

If $T>T_F$ ($r_0>0$) and $\eta>\eta_c$ the global minimum corresponds to the modulated phase when $r_0+2e_{min}/\delta<0$, i.e. $T > T_F -2 e_{min}/a\delta$. Therefore, there is a transition line at $T_c(\eta) = T_F -2 e_{min}(\eta)/a\delta$. The order parameter changes continuously at $T_c$  ($M^2 =-(r_0+2e_{min})=0$), but $s$ changes discontinuously. In Fig.\ref{pd1} we illustrate the typical topology of  the phase diagram for the particular case $\delta=6$. All the solid lines in \ref{pd1} correspond to second order phase transitions. We also show the crossover line between the region where the magnetization profile shows a significative canting angle (canted stripes) and the region where the local magnetization is almost perpendicular to the plane (Perpendicular stripes).

Although the paramagnetic solution is not a global minimum of $f$ when $\eta>\eta_c$ and $T>T_F$, it could be still a local minimum provided that $r_0+G<0$ for some  values of $s$ and $k_0$. From Eq.(\ref{stabM0}), such condition ensures the local stability against variations of $M$. However, since all the rest of the second derivatives cancel, the complete stability of the paramagnetic solution is beyond the linear analysis.  We verified numerically that indeed the paramagnetic phase remains locally stable (metastable) below  $T=T_F$ at fixed $\eta>\eta_c$. Such metastability is a result of the high degeneracy of the paramagnetic solution under the present approximation, so it appears to be a spurious result. However, it can be indicative of a change in the order of the transition if the approximation is improved. Indeed, there are several evidences towards the first order nature of the stripes-disordered phase transition \cite{CaStTa2004,CaMiStTa2006,CaBiPiCaStTa2008}.

\section{Monte Carlo simulations}
\label{mc}

In order to compare the mean field results with the behavior of a specific microscopic model, we performed Monte Carlo (MC)
simulations using a Heisenberg model with exchange and dipolar interactions, as well as uniaxial out of plane anisotropy.
The model, which describes an ultrathin magnetic film (see Ref.\onlinecite{PiBiStCa2010} and references therein) can be
characterized by the dimensionless Hamiltonian:
\begin{widetext}
\begin{equation}
{\cal H} = -\delta \sum_{<i,j>} \vec{S}_i \cdot \vec{S}_j +
\sum_{(i,j)} \left[ \frac{\vec{S}_i \cdot \vec{S}_j }{r_{ij}^3} - 3 \,
\frac{(\vec{S}_i \cdot \vec{r}_{ij}) \; (\vec{S}_j \cdot \vec{r}_{ij})}{r_{ij}^5} \right]
- \eta \sum_{i} (S_i^z)^2,
\label{hamiltoniano}
\end{equation}
\end{widetext}
\noindent where the exchange and anisotropy constants are normalized relative to the dipolar coupling
constant, $<i,j>$ stands for a sum over nearest neighbors pairs of sites in a square lattice with $N = L_x \times L_y$ sites (the lattice parameter is taken equal to one),
$(i,j)$ stands for a sum over {\it all distinct} pairs and $r_{ij}\equiv |\vec{r}_i - \vec{r}_j|$ is the
distance between spins $i$ and $j$. Each spin is defined by a unit vector with components $S^x,S^y,S^z$. All the simulations were done using the Metropolis algorithm, and periodic boundary conditions
were imposed on the lattice by means of the Ewald sums technique. We focus our simulations on  the case  $\delta=6$,
where the system presents a canted equilibrium state at zero temperature for a wide range of the anisotropy values\cite{PiBiStCa2010}.

 The phase diagram was
obtained  by measuring the out-of-plane magnetization;
\begin{equation}
 M_z \equiv \frac{1}{N}\sum_{\vec{r}} \left< S^z(\vec{r}) \right>,
\label{mz}
\end{equation}
(the in plane components are defined in a similar way) the in-plane magnetization;
\begin{equation}
 M_{||} \equiv \sqrt{(M_x)^2 + (M_y)^2},
\end{equation}
and an orientational order parameter\cite{CaBiPiCaStTa2008};
\begin{equation}
 O_{hv} \equiv \left< \left| \frac{n_h-n_v}{n_h+n_v} \right| \right>,
\end{equation}
\noindent where  $\left< \cdots \right>$ stands for a thermal average, $n_h$ ($n_v$) is the number of horizontal
(vertical)  pairs of nearest neighbor spins with antialigned perpendicular component, {\it i.e.},
\begin{equation}
 n_h = \frac{1}{2}\sum_{\vec{r}} \,\left\{1-sig\left[S^z(r_x,r_y),\,S^z(r_x+1,r_y)\right]\right\}
\end{equation}
\noindent and a similar definition for $n_v$, where $sig(x,y)$ is the sign of the product of $x$ and
$y$. To obtain the stripe width of the modulated states we considered the structure factor $|\hat{S}(\vec{k})|^2$, where

\begin{equation}\label{struct}
    \hat{S}(\vec{k}) = \frac{1}{\sqrt{N}}\, \sum_{\vec{r}} S^z(\vec{r})\, e^{-i\vec{k}.\vec{r}}.
\end{equation}

\noindent The stripe width was calculated using the expression
\begin{equation}\label{struct}
h=\pi/k_{max},
\end{equation}

\noindent where $k_{max}$ is the modulus of the wave vector that maximize $|\hat{S}(\vec{k})|^2$.

In order to find the equilibrium phase diagram, we carried out the simulations with two protocols
for the independent parameter (temperature, anisotropy or external field).
In the first protocol, we varied the independent parameter linearly with the simulation time,
increasing or decreasing it at a given rate $r$,  keeping the rest of the parameters fixed.
For instance, if we choose the temperature, then $T(t)=T(0) + r\,t$ where $t$ is the simulation
time measured in units of Monte Carlo Steps (MCS). Each MCS corresponds to $N$ single spin updates
of the Metropolis algorithm. The initial spin configuration at $T(0)$ was previously obtained by
performing $t_e$ MCS to equilibrate. The order parameters were calculated along the simulation
and averaged over many realizations to improve statistics. We called this protocol
``linear variation of parameters'' (LVP). The second one was a ladder protocol. For instance, in the case of $T$ being the independent parameter,
the system is initialized at the paramagnetic state at a high
temperature, and then temperature is reduced at discrete steps.
The initial configuration for each temperature is the last one of the
previous step. At each step we discarded the first $t_e$ MCS in order to equilibrate, then
we calculated the averages over the next $t_m$ MCS.

\begin{figure}
  \includegraphics[angle=0,scale=0.39]{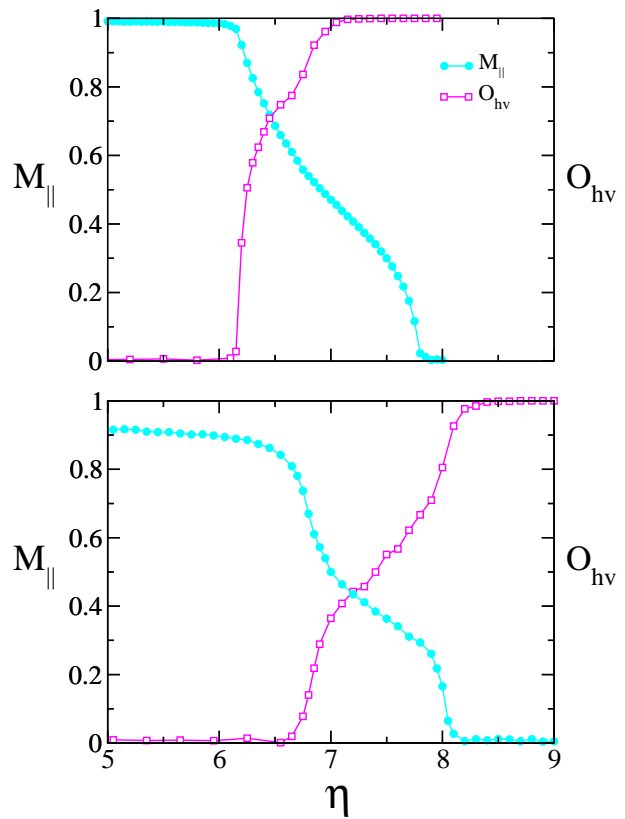}
\begin{center}
\caption{\label{mc1} (Color online) In plane magnetization and orientational order parameter as a
function of anisotropy for $\delta=6$. $T=0.1$ (upper panel) and $T=1.0$ (lower panel).
}
\end{center}
\end{figure}

First, we calculated $M_{||}$ and $O_{hv}$ as a function of the anisotropy for two fixed temperatures.
We applied the LVP protocol to increase $\eta$ from a small value $\eta<\eta_c$, starting from an equilibrated  in-plane ferromagnetic configuration.
In these simulations we used $L_x=L_y=L$, with $L=80$ and $120$.
The parameter variation rate $r$ was ranged from $r=10^{-5}$ to $10^{-7}$, depending on the
temperature and the system size.
The typical  behavior of the order parameters at low temperatures ($T = 0.1$ and $T=1.0$)
is illustrated in Fig.\ref{mc1}. Three different behaviors can be identified:
the planar ferromagnetic state at low anisotropies, characterized  by $M_{||} \neq 0$ and $O_{hv}=0$;
a canted striped state at intermediate values of $\eta$, with both $M_{||} \neq 0$ and $O_{hv}\neq 0$;
and a perpendicular striped state at large enough values of $\eta$, characterized by
$M_{||} = 0$ and $O_{hv}\neq 0$.
At each temperature the transition points are identified with the values at which the corresponding order
parameter becomes zero.
%In particular, at high temperatures $T > 2.5$ there is no orientational order, namely,
%the orientational parameter $O_{hv}=0$ for any value of $\eta$ and the system undergoes a direct
%transition from a planar ferromagnet into a paramagnetic state with $M_{||} = 0$ as $\eta$ increases.

\begin{figure}
\begin{center}
 \includegraphics[scale=0.28,angle=-90]{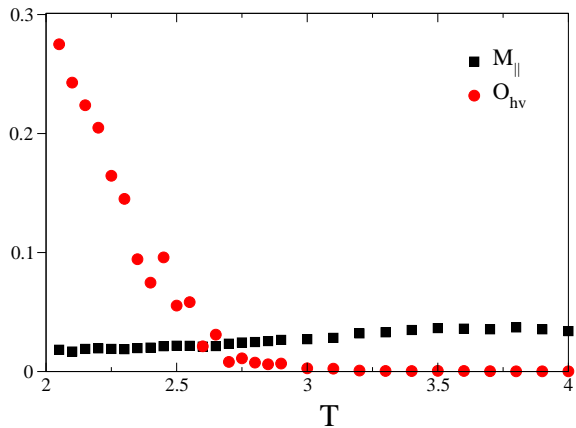}
\caption{\label{mc2} (Color online) Orientational order parameter (red circles) and  in-plane magnetization (black squares)  as a function of temperature for $\delta=6$ and $\eta=8.7$.}
\end{center}
\end{figure}

At large anisotropy values ($\eta > 8.4$), the in-plane ferromagnetic phase is absent for any value of the temperature.
The system  undergoes a direct transition from an almost perpendicular striped state into the
paramagnetic state as the temperature is increased. This can be seen in Fig.\ref{mc2},
where the order parameters were computed using a ladder protocol cooling from $T=4.0$,
with $L=120$, $t_e=10^5$, and $t_m=10^5$.

\begin{figure}
  \centering
  \includegraphics[scale=0.29]{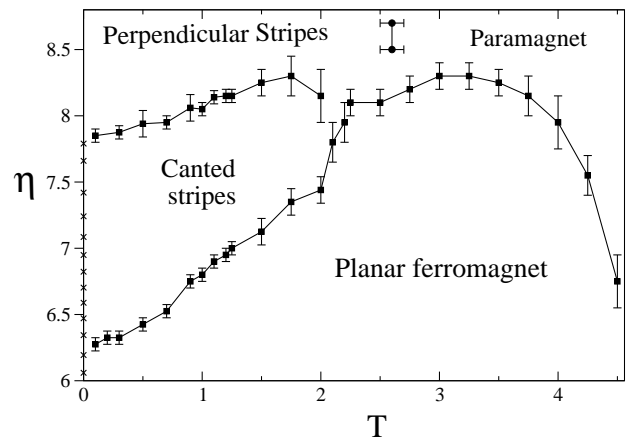}
  \caption{\label{mc6} Monte Carlo phase diagram  for $\delta=6$.
}
\end{figure}

By means of the methods described above, we obtained the phase diagram
shown in Fig.\ref{mc6}. Although, the global topology of this diagram is in agreement
with that obtained by the mean field theory (see Fig.\ref{pd1}), some noticeable
differences exist, which appear to be an artifact of the mean field approach.
Some of them will be discussed in section \ref{conc}.
The presence of a canted striped region is in agreement with previous
MC calculations carried out by  Whitehead {\it et al.} in Ref. \onlinecite{WhMaDe2008}
for $\delta=4.5$ and with the zero temperature behavior of the model\cite{PiBiStCa2010}.
In particular, our results show that the canted region is larger than the
corresponding to $\delta=4.5$.

\begin{figure}
\begin{center}
\includegraphics[scale=0.28]{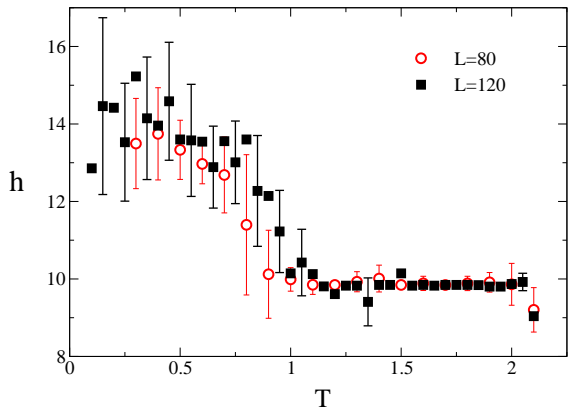}
\caption{\label{mc3} (Color online) Stripe width as a function of temperature for $\delta=6$ and $\eta=7.5$.  The simulation parameters were $L=80$, $r=-10^{-6}$, $t_e=10^{5}$ and $t_H=10^{5}$ (circle)
and $L=120$, $r-=10^{-7}$, $t_e=10^{6}$ and $t_H=10^{6}$ (box). The error bars are the
standard deviation taken into account many realizations, some of them are omitted for clarity.}
\end{center}
\end{figure}

We explored the canted region looking at the variation of the stripes width $h$.
This is a difficult task because the stripe width variation is
mediated by the formation of topological defects (usually stripe dislocations) which need long
simulation times to nucleate and move. An acceleration of this process
was obtained when we added an in-plane external field term to Eq.\eqref{hamiltoniano}
of the form $- \xi \sum_{i} (S_i^x)$ and applied a LPV protocol with $\xi$ as the free
parameter. The value of $\xi$ vary from $\xi(0)=H_x$ where $H_x$ is an external magnetic field
 strong enough to saturate the magnetization in the x
direction, to zero (zero-field condition). After several tests, we
found that $H_x=0.5$ was optimal for all the regions of the phase diagram
studied in this work.
Then,  $t_H$ extra MCS were performed before calculating $h$. The stripe
width  shown as a function of temperature for $\eta=7.5$ in Fig.\ref{mc3} is the
result of an average performed over several realizations of the LPV
protocol and the error bars correspond to the dispersion of $h$.

The stripe width at the SRT is $h=10$ and remains constant down to $T=0.75$, where it displays
a steep increase. The widest stripes are observed at low temperatures reaching a maximum
value  of $h \sim 14$, close to the zero temperature value, $h=17$, calculated
previously\cite{PiBiStCa2010}. The $h$ values obtained for both system sizes
are almost undistinguishable, showing that finite size effects are negligible.

\begin{figure}
  \centering
\includegraphics[scale=0.9]{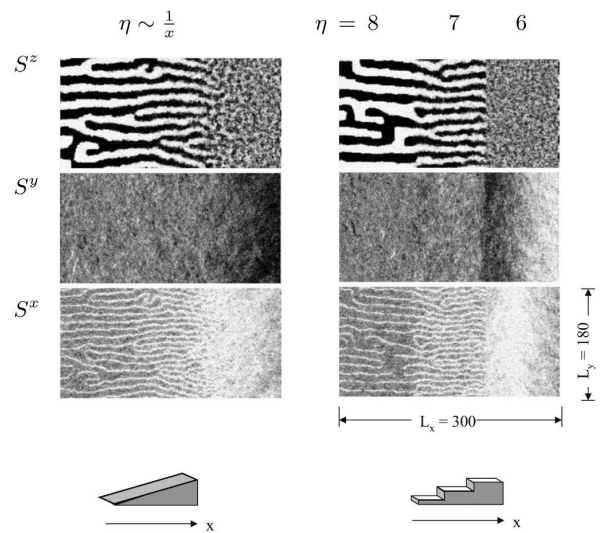}
\caption{\label{mc4} Snapshots of the spins components for two wedges in equilibrium for $\delta=6$ at $T=1$.
The film width functional dependence
varies as a linear function of $x$ (left) or as equal-spaced steps of values
$\eta=8,7,6$ (right), as schematized at the bottom of the columns. The simulation parameters are $r=10^{-5}$ and $t_e=10^5$. }
\end{figure}

\begin{figure}
\begin{center}
\includegraphics[scale=0.28]{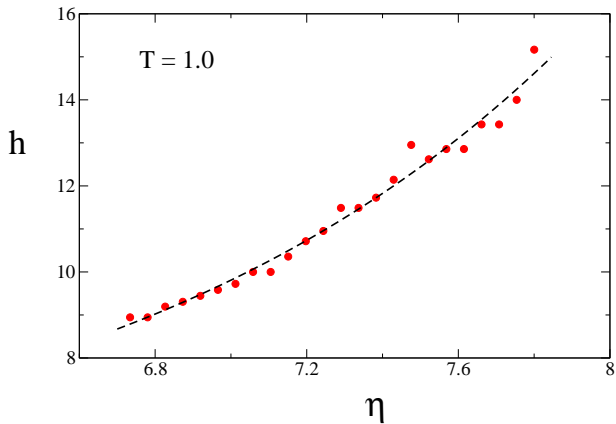}
\caption{\label{hvseta} Stripe width as a function of $\eta$ corresponding to the wedge like film of the left
panel of Fig.\ref{mc4} ($\delta=6$). The dashed line correspond to a parabolic fiting.}
\end{center}
\end{figure}

We next analyzed the stripe width variation with the anisotropy, which is closely related to the variation with the film thickness $d$. Indeed, previous numerical simulations suggest that the effective out of plane anisotropy varies inversely with the film thichness\cite{CaBiPiCaStTa2008} $\eta \sim 1/d$. A usual experimental technique to analyze the effects of the film thickness on the magnetic patterns is to take images on wedge-like ultrathin films, where the film width
is a function one of the axis (eg. Refs.\onlinecite{VaStMaPiPoPe2000,WuWoSc2004}).
We modeled these systems assuming  the relation $\eta = 1/d(x)$,
where $d(x)$ is the local film width that depends on the $x$ position.
The simulations were performed over rectangular lattices of size $L_x=360, L_y=180$  using
the external field protocol used in the study of the stripe width variation and periodic boundary conditions in the $y$ direction.
The results are shown in Fig.\ref{mc4} in two columns, each column being related to different
$d(x)$ assumptions. The left one corresponds to a continuous linear variation of $d(x)=a+bx$, where $a$ and $b$ were chosen such that $\eta$ varies from $\eta=6$ at $x=1$
to $\eta=8$ at $x=300$. The right column corresponds to a ladder structure, where $d(x)$ varies at equal-spaced steps corresponding to  the values
$\eta=8,7,6$ from left to right. The figure presents typical snapshots of equilibrated magnetic patterns at fixed temperature.
Each pixel represents a spin component in gray scale, ranging from white when the value
is $1$ to black when it is $-1$. From the $S^z$ component behavior turns out that the stripes width decreases
as $x$ increases until the SRT. Once the SRT is reached, the spins
are ferromagnetically ordered in the same direction of the in-plane component of the
magnetization in the walls.
Moreover, the $S^x$ components in the walls are along the stripes direction,
showing  the they are Bloch's walls as expected\cite{Po1998}.
The stripe width reduction occurs by the insertion of new stripes from the low anisotropy (higher thickness) region, in agreement with experimental results on Fe on Cu\cite{PoVaPe2003} and Fe/Ni on Cu\cite{WuWoSc2004} films. These results give further support to the assumption
$\eta\sim 1/d$ and suggest that the observed stripe width variation  with the film thickness is due to the induced anisotropy gradient.
In fact, the $\eta \sim 1/d$ dependency is probably related to the contribution to the effective anisotropy coming from the short
range part of the dipolar energy, which can be assumed proportional to the film thickness\cite{KaPo1993} (at least in the ultrathin
limit).
The stripe width variation with $\eta$ in the wedge like film of Fig. \ref{mc4} is shown in Fig. \ref{hvseta}.
We also performed a series of simulations on lattices with $L_x=L_y$ and uniform anisotropy, for different values of $\eta$.
The equilibrium average stripe width agreed with that observed in the wedges.

We verified that the increase in the stripe width as $\eta$ increases follows a series of steps in a similar way to that observed at zero temperature\cite{PiBiStCa2010}. In other words, the canted region in the phase diagram of Fig.\ref{mc6} is composed by a series of transition lines (not shown for clarity) that follow a similar direction as the SRT line and converge to the zero temperature transition points between different stripe width ground states\cite{PiBiStCa2010}. Every time the anisotropy crosses one of such lines the stripe width increases by one unit (dynamically mediated by defects). In this way, the stripe width variation with temperature (horizontally crossing of those lines in Fig.\ref{mc6}) is related to the ground state structure of the system. In this context, the absence of stripe width variation with temperature for $\eta > 8.5$ is related to the fact that for $\delta=6$ the ground state stripe width has already saturated\cite{PiBiStCa2010}.

%Due to the periodic boundary conditions, a slight rotation of the in-plane spins from
%the $x$ to the $y$ direction is noticed from $S^x$ and $S^y$ snapshots near the $(x=360,y)$ border.
%

\begin{figure}
  \centering
 \includegraphics[scale=0.28]{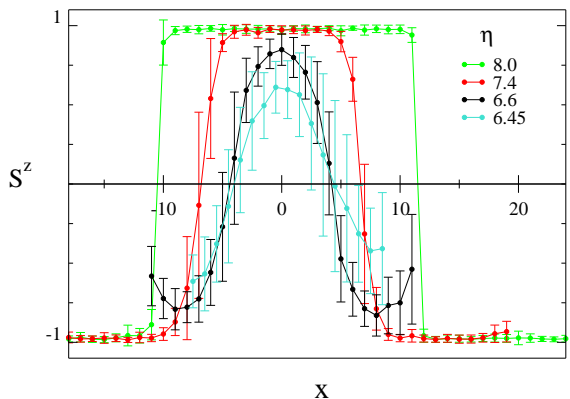}
  \caption{\label{mc5} (Color online) Mean stripe magnetization profiles in the perpendicular direction for $\delta=6$ at $T=0.5$.
Each curve is an average
of consecutive profiles along a stripe and the error bars correspond to the standard deviation.}
\end{figure}

Finally, we analyzed the magnetization profile variation of the stripe pattern as a function of anisotropy, which can be seen in Fig.\ref{mc5}.
These results were obtained by averaging  over $25$ adjacent profiles $S^z(x,y_0)$ in a system with uniform anisotropy
in equilibrium at $T=0.5$. Topological defects like dislocations were avoided in the calculation.
We thermalized the system using the external field protocol with parameters $L_x=L_y=144$, $r=10^{-5}$, $t_e=10^5$ and $t_m=10^4$.
At $\eta=6.45$ the system is in the canted state, the walls are wide with sinusoidal like shape.
The perpendicular components of the spins at the center of the stripes is lower than $1$,
meaning that all the spins  have an in-plane component aligned with the stripes, thus contributing
to the in-plane magnetization as pointed out in Fig.\ref{mc1}.
As the anisotropy increases, the spins at the center of the stripes become perpendicular
to the plane and the walls narrow but are still extended. The same behavior of the magnetization profile is observed at zero temperature \cite{PiBiStCa2010}.
Finally, at high enough anisotropy values ($\eta\geq 8.0$), the stripes widen and the wall widths $w$ become close to one.

It is worth to note that the fluctuations of the spin directions at low and intermediate anisotropy values are stronger within the
walls, as can be observed from the error bars. This suggests that these spins are less restricted
to move  and therefore facilitate defects mobility. This  explains the higher efficiency of the previously used in-plane field protocol to obtain thermal equilibration, through the interaction between the external field and the large in plane components inside the walls.

 Interestingly, the change in the magnetization stripe profile as the anisotropy increases closely resembles that observed as the temperature decreases, both experimentally\cite{ViSaPoPePo2008} and in mean field theories\cite{ViSaPoPePo2008,CaCaBiSt2011}.

\section{Discussion}
\label{conc}

Our mean field results suggest that the global topology of the phase diagram observed  numerically  for low values of $\delta$ (both from previous\cite{CaBiPiCaStTa2008,WhMaDe2008} and from the present simulations) is robust, at least under the validity conditions of the present approximation, namely, in the low anisotropy region close to the SRT. For large enough values of the anisotropy the approximation breaks down, as evidenced by the unphysical monotonous increase of the transition temperature between the stripes and paramagnetic phases in the large $\eta$ region. This breakdown is on the basis of the present MF approximation, namely, in the effective free energy Eq.(\ref{Hreal}). Such free energy can be obtained variationally from a partition function ${\cal Z} = {\rm Tr} e^{-\beta H[{\bf M}]}$, where the coarse-grained Hamiltonian $H[{\bf M}]$ has the same structure of\cite{Goldenfeld} Eq.(\ref{Hreal}). The effective free energy $F[{\bf M}]$ is then the order zero term in an expansion of  $H[{\bf M}]$ around its minimum when fluctuations are neglected. If the $\eta\to\infty$ limit is taken {\it a priori} of such expansion, all the configurations with non zero in plane magnetization components get zero statistical weight and a different Landau Ginzburg free energy (which depends only on the scalar field $M^z$) is obtained\cite{CaCaBiSt2011}. Therefore, the correct stripe-paramagnet critical temperature must converge to the ($\eta$-independent) value predicted by the last free energy when $\eta\gg 1$. In other words, even within the mean field theory the correct behavior cannot be obtained as the $\eta\to\infty$ limit of the present approach.

While the previous difference (vertical line vs. finite slope) between the MC and the MF phase diagrams is particular of the present approach, some others appear to be associated to general features of the mean field theory. For instance, the transition line between the planar ferromagnet and the canted-stripe phases computed within the MF approximation is horizontal, while it shows a finite slope when extracted from MC simulations. This seems to be a direct effect of neglecting thermal fluctuations, since theoretical works show that those fluctuations renormalize the dipolar and
anisotropy coupling parameters in such a way that the anisotropy
$K(T)$ diminishes faster than the dipolar coupling
constant\cite{PePo1990,PoRePi1993} $g(T)$ (in our notation,$\eta=K/g$). Those works
predict a linear dependence of the reorientation transition temperature
 with anisotropy with positive slope, which is
roughly in agreement with the transition lines obtained from MC
simulations.
Finally, the transition line between the planar ferromagnet and the paramagnetic phases is a vertical straight line
in MF diagram while it shows some slope in the MC diagram. This is because of the simplifying assumption that the (coarse grained) phenomenological transition temperature $T_F$ is independent
of $\eta$. While the tendency of the transition line in the MC diagram suggests  that this assumption may appropriately describe the very
low $\eta$ limit, it clearly fails for large enough values of $\eta$. An increase in the perpendicular anisotropy should destabilize
the planar ferromagnetic phase, thus decreasing the critical temperature.

One fact that emerges, both from  our mean field and Monte Carlo results, is the strong influence of the ground state properties on the finite temperature behavior close to the SRT. One example is the presence of canted states close to the SRT line. Comparing with previous MC results for\cite{WhMaDe2008} $\delta=4.5$, the phase diagram canted region becomes wider for $\delta=6$, consistently with the zero temperature phase diagram\cite{PiBiStCa2010}. However,  the range of values of $\eta$ where the ground state canted angle is different from zero becomes extremely narrow as $\delta$ is further increased. Hence, our mean field results suggest that those states would be present at finite temperature only very close to the SRT line for any realistic value of $\delta$. Another example is the stripe width variation and the magnetization stripes profile change with $\eta$, that closely follow the zero temperature behavior\cite{PiBiStCa2010}.
Moreover, the qualitative agreement between our MC simulations on wedges and experimental results on Fe/Ni
ultrathin films\cite{WuWoSc2004} supports inverse relationship between out of plane anisotropy and film thickness $\eta\sim 1/d$.

The  correlation between the stripe width variation with the temperature and with the anisotropy observed close to the SRT in the present simulations is another interesting fact. As previously pointed out\cite{PoVaPe2003}, varying the film thickness  (always in the ultrathin limit) produces a similar effect as changing the temperature, thus leading to an
``inverse effective temperature'' interpretation of the thickness\cite{PoVaPe2003}. Considering the relation  $\eta\sim 1/d$, this appears to be consistent with the similarity observed between the change in the magnetization profile when $\eta$ is varied and that observed in Fe films when the temperature is varied\cite{ViSaPoPePo2008}. Such set of similarities suggest that a deeper analysis about the interplay between temperature and anisotropy could shed an additional light about the origin of the strong stripe width variation with temperature observed in ultrathin magnetic films.

This work was partially supported by grants from
 CONICET (Argentina) and SeCyT, Universidad Nacional de C\'ordoba
(Argentina).

\bibliographystyle{apsrev}
%\bibliography{ultrathin}

\end{document}